\documentclass[12pt]{article}
\usepackage{amssymb,graphicx,epsfig,cite}
\def\baselinestretch{1.25}
\def\beq{\begin{eqnarray}}
\def\enq{\end{eqnarray}}
\newcommand{\lsim}{\raisebox{-0.13cm}{~\shortstack{$<$ \\[-0.07cm] $\sim$}}~}
\newcommand{\gsim}{\raisebox{-0.13cm}{~\shortstack{$>$ \\[-0.07cm] $\sim$}}~}
\def\lapp{\mathrel{\rlap{\raise.5ex\hbox{$<$}}{\lower.5ex\hbox{$\sim$}}}}
\def\gapp{\mathrel{\rlap{\raise.5ex\hbox{$>$}}{\lower.5ex\hbox{$\sim$}}}}
\begin{document}
\thispagestyle{empty}
\begin{flushright}
IPMU12-0068          \hfill                  TIFR-TH/12-12 
\end{flushright}

\vspace*{-0.2in}
\begin{center}

{\Large\bf Using Jet Substructure at the LHC to Search for the \\ [1mm] 
Light Higgs Bosons of the CP-Violating MSSM }

\bigskip

{\sl Biplob Bhattacherjee $^{a,}$}\footnote{\sf E-mail: 
biplob.bhattacherjee@ipmu.jp},
{\sl Amit Chakraborty $^{b,}$}\footnote{\sf E-mail: tpac@iacs.res.in}, 
\\
{\sl Dilip Kumar Ghosh $^{b,}$}\footnote{\sf E-mail: tpdkg@iacs.res.in}, 
and
{\sl Sreerup Raychaudhuri $^{c,}$}\footnote{\sf E-mail: 
sreerup@theory.tifr.res.in}

{\small
$^a$ Kavli Institute for the Physics and Mathematics of the Universe 
(WPI), \\ The University of Tokyo, Kashiwa, Chiba 277-8583, Japan.

$^b$ Department of Theoretical Physics, Indian Association for the 
Cultivation of Science, \\ [-1mm] 2A \& 2B, Raja S.C.\,Mullick Road, 
Jadavpur, Kolkata 700\,032, India.

$^c$ Department of Theoretical Physics, Tata Institute of Fundamental 
Research, \\ [-1mm] 1, Dr. Homi Bhabha Road, Mumbai 400\,005, India. }

\bigskip

{\large\bf ABSTRACT}
\end{center}

\vspace*{-0.4in}
\begin{quotation}
\small
\baselineskip 14pt
\noindent The $CP$-violating version of the Minimal Supersymmetric 
Standard Model (MSSM) is an example of a model where experimental data 
do not preclude the presence of light Higgs bosons in the range around 
10 -- 110 GeV.  Such light Higgs bosons, decaying almost wholly to 
$b\bar{b}$ pairs, may be copiously produced at the LHC but would remain 
inaccessible to conventional Higgs searches because of intractable QCD 
backgrounds. We demonstrate that a significant number of these light 
Higgs bosons would be boosted strongly enough for the pair of daughter 
$b$-jet pairs to appear as a single `fat' jet with substructure. Tagging 
such jets could extend the discovery potential at the LHC into the 
hitherto-inaccessible region for light Higgs bosons.
\baselineskip 18pt
\end{quotation}
\normalsize
\centerline{PACS Nos:  12.60.Fr, 14.80.Da, 13.85.Qk}

\vfill


\newpage
\section{Introduction: Survival of Light Higgs Bosons}

\vspace*{-0.2in} After four decades of relative stagnation, the physics 
of elementary particles has again reached a stage when empirical 
discoveries could lead the day. Theoretical structures built with much 
care and ingenuity over the previous decades now lie in serious danger 
of collapsing under the onslaught of data from the Large Hadron Collider 
(LHC). The most important of these theoretical structures under test is, 
of course, the Standard Model (SM) of particle physics. The first -- 
and, it turns out, most stringent -- test of the SM lies in the 
discovery of the predicted \cite{pwhiggs,goldstone,hhg,lepwwg} Higgs 
boson $H^0$. It is now universally known that a discovery has been made 
\cite{Higgs_discovery} of a boson of mass around 125~GeV which seems to 
resemble the Higgs boson of the SM \cite{bestfit}, but further precise 
measurement of its couplings are required before it can be established 
as the SM Higgs boson in a manner convincing to all. The situation is 
expected to become much more clear at the end of the current year, after 
more data is collected and analysed.

Whatever be the outcome of these Higgs boson measurements, the issue of 
light Higgs bosons (i.e. masses $\lsim 100$~GeV) will still remain wide 
open, so far as the LHC is concerned. This is because signals for a 
light Higgs boson decaying principally to $b$-jets will be completely 
swamped by the QCD background for production of $b\bar{b}$ pairs (and 
other dijets) -- which is to say that such light Higgs bosons will be, 
for all practical purposes, invisible at the LHC. At a first reading, 
the above argument might seem to be purely academic, in view of the 
definite bound of $M_H \gsim 114.4$ GeV reported by the LEP-2 
electroweak working group \cite{lep2bound}. This is certainly true of 
the SM Higgs boson. However, we must remember that this bound arises 
from the negative results of searches for the Higgstrahlung process, 
viz.,
\begin{equation}
e^+ + e^- \to Z^\ast \to Z^0 + H^0  \to (\ell^+\ell^-) + (b \bar{b}) 
\end{equation}
where $\ell = e, \mu$ or $\tau$, and is, therefore, strongly dependent 
on the $HZZ$ coupling. This coupling is, of course, fixed in the SM, and 
therefore the lower bound of 114.4~GeV is also fixed in the SM. On the 
other hand, any model with new physics beyond the SM (BSM) which can 
accommodate a significantly smaller $HZZ$ coupling will immediately 
evade this bound. Thus, we can even now ask the question whether the 
observed new particle is the only Higgs boson, or is it one of a pack of 
scalars in some BSM physics model, where the others are invisible at the 
LHC because they are light?

Whenever new physics beyond the SM is indicated, the usual model of 
choice is the Minimal Supersymmetric Standard Model (MSSM) or one of its 
many variations. Even in the MSSM with all real and CP-conserving 
parameters, the lower bound on the lightest Higgs boson mass (93~GeV for 
$\tan\beta > 6$) is different from that of the SM Higgs mass bound 
\cite{lep2mssm}. Not surprisingly, even this bound can be relaxed 
considerably in the presence of CP violation in the Higgs sector 
\cite{gunion_haber}, where the lightest Higgs boson $h^0$ can have a 
substantial CP-odd component -- which immediately implies a highly 
suppressed $h^0ZZ $ coupling (see below).

Though an overwhelming majority of studies of the MSSM have assumed that 
the supersymmetric parameters are real and $CP$ conserving, it may be 
recalled that of the 105 new undetermined parameters in the MSSM, only 
62 of them are CP-conserving, while as many as 43 are CP-violating. Many 
of these CP-violating phases cannot be rotated away by simple 
re-definition of fields and this is known to lead to new sources of CP 
violation which may come in useful to explain the observed level of 
baryon asymmetry in the Universe \cite{Dine:2003ax}. However, these 
additional phases, especially those involving the first two generations, 
lead to large electric dipole moments (EDMs) of $e^\pm$ and $\nu_e$ as 
well as of mercury (Hg) atoms -- which come in conflict with the 
experimentally measured upper bounds on these EDMs \cite{edm1, edm2, 
edm3}. The new CP phases may, therefore, be expected to be severely 
constrained. Fortunately, it turns out that such constraints are 
strongly dependent on the CP-conserving model parameters -- specifically 
the nature of the mass spectrum involved in the calculation of those 
EDMs -- and it has been explicitly shown that substantial cancellations 
among different EDM diagrams end up in allowing some combinations of the 
CP-violating phases to be large \cite{edm4}. For example, the EDM 
constraints require the phase $\phi_\mu$ of the higgsino mass parameter 
$\mu=| \mu|e^{i\phi_\mu}$ to be generally constrained to 
$\phi_\mu^{}\lsim 10^{-2}$ (unless we set the masses of all the 
sfermions very high). But if the sfermions of the first two generations 
are of the order of a few TeV \cite{edm5} and we do not assume 
universality of the trilinear scalar couplings $A_f$ \cite{edm6,edm7}, 
then $\phi_\mu$ can be considerably larger even if we keep the sfermions 
of the third generation light enough for easy detection at the LHC. The 
presence of CP-violating phases can substantially modify Higgs boson and 
superparticle production at colliders as well as decay modes and this 
has been the subject of several investigations \cite{CPv-susyhiggs} in 
the context of collider signals.

Though the new CP-violating phases appear only in the soft 
supersymmetry-breaking parameters $\mu$, $A_f$ and the three gaugino 
masses $M_1, M_2$ and $M_3$, some of them can induce CP violation at the 
one-loop level in the Higgs potential even if the tree-level Higgs 
potential is CP-conserving \cite{CPv1, CPv2, CPv3, CPv4, CPv5, CPv6}. 
The quadratic terms for neutral states in this one loop-corrected Higgs 
potential can be written in terms of a $3\times 3$ mass-squared matrix 
${\cal M}_{ab}^2$, where non-zero off-diagonal terms ($a \neq b$) 
involve mixing between scalar and pseudoscalar states. This is unlike 
the CP-conserving MSSM, where the scalar states ($h^0, H^0$) and the 
pseudoscalar state $A^0$ do not mix. After diagonalisation, the physical 
neutral Higgs states $h^0_1$, $h^0_2$ and $h^0_3$ (in ascending order of 
mass) become admixtures of CP-even and CP-odd states. Since the 
pseudoscalar states do not couple to $ZZ$ pairs, it is obvious that the 
$h_i^0ZZ$ couplings ($i = 1,2,3$) arise only from the CP-conserving 
components of $h^0_i$ and therefore will be suppressed by the 
corresponding mixing angles. There exists enough freedom in choice of 
parameters in this CP-violating MSSM for at least some of these $h^0_i$ 
states to be light --- which would make them invisible to conventional 
searches at the LHC. In fact, a set of benchmark points have been 
defined to showcase the maximal effect of CP violation in the MSSM Higgs 
sector, and this set goes by the name: `CPX scenario' \cite{CPv5}.

The LEP collaborations have searched for the processes
$$
e^+e^- \to \left\{ \begin{array}{l}
Z^\ast \to Z^0 + h^0_i \to \ell^+\ell^-+ b\bar{b} ~~~ (i = 1,2,3)\\
h^0_1 + h^0_2 \to 4b \\
h^0_1 + h^0_2 \to 3h^0_1 \to 6b \\
Z^0 + h^0_2 \to Z^0 + 2h^0_1 \to  (\ell^+\ell^-) + 4b 
\end{array} \right.
$$
in the CP-violating MSSM Higgs sector based on the CPX scenario 
\cite{LEPsearch}. For certain choices of CP-violating parameters within 
the CPX scenario, the LEP-2 data allow for a much lighter Higgs boson 
$h^0_1$ with a mass $M(h_1) \approx 40$--$50$ GeV 
\cite{lep2mssm,carena,Abbiendi:2004ww} because, as expected, there are 
very substantial reductions in the $h^0_1 ZZ$ coupling \cite{CPv5}. It 
turns out that for the selfsame sets of parameters the couplings 
$h^0_1WW$, $h^0_1ZZ$ and $ h^0_1t\bar t$ all get reduced simultaneously, 
as a result of which none of the canonical search channels for $h^0_1$ 
at the Tevatron and LHC are expected to be viable \cite{carena, 
Buescher:2005re, Schumacher:2004da, Accomando:2006ga}. This implies that 
there is a `blind spot' or `hole' in the parameter space which is 
permitted by all experimental data till date. Different search 
strategies in the future runs of the LHC have been proposed to close 
this `blind spot' \cite{Buescher:2005re,Ghosh:2004cc,Ghosh:2004wr, 
Bandyopadhyay:2007CP}. These have varying levels of success, depending 
on the specific choice of parameters, but none can be said to be the 
definitive search strategy for light Higgs bosons belonging to this 
inaccessible region.

In this work, we explore a method of searching for these light Higgs 
bosons which is based essentially on kinematics, and hence is not overly 
sensitive to the parameter choices of the theory. Our strategy is to 
apply to the case of the light Higgs bosons of the CP-violating MSSM 
some of the techniques developed recently for tagging a heavy boosted 
particle decaying to a single fat jet with substructure 
\cite{Seymour,Butterworth:2007ke, BDRS,Thaler:2008ju,Kaplan:2008ie,rev}. 
Since the $h_1^0$, $h^0_2$ and $h^0_3$ can all decay to a pair of 
highly-boosted $b$-jets, one could ask if this technique can suitably 
identify the Higgs boson states over and above the enormous QCD 
background at the LHC. Daunting as the task may seem at first, we 
demonstrate that this technique, in fact, works quite well, and can be 
considered as an important probe of the CP-violating Higgs sector. The 
basic technique can, in fact, be used in the wider context of light 
scalar states, such as those considered recently in 
Ref.~\cite{GhoGuSen}.

\vspace*{-0.2in}
\section{The Higgs Sector of the CP-violating MSSM}

\vspace*{-0.2in} As already mentioned in the introductory section, the 
non-vanishing phases of $\mu$ and/or the trilinear scalar couplings 
$A_t$ and $A_b$ can induce explicit CP violation in the Higgs sector. 
Since the only Yukawa interactions of the Higgs bosons which can have 
any significant effects are those to top and bottom squarks, this 
feature is reflected in the trilinear couplings as well, and thus the 
only relevant CP phases are $\phi_\mu = {\rm Arg}[\mu]$, $\phi_t = {\rm 
Arg}[A_t]$ and $\phi_b = {\rm Arg}[A_b]$. Given these, the scalar 
potential, even though invariant under CP-transformation at tree level, 
receives CP-violating contributions through one-loop corrections.

We briefly review the formalism required to include CP-violating effects 
in the MSSM. As is usual, we write the two scalar doublets as gauge 
eigenstates in the form
\begin{equation}
\Phi_1 = 
\left( \begin{array}{c} \phi_1^+ \\ \phi_1^0 + i \eta_1^0 \end{array} \right)
\qquad\qquad
\Phi_2 = 
\left( \begin{array}{c} \phi_2^0 + i \eta_2^0 \\ \phi_2^- \end{array} \right)
\end{equation}

Once we allow the parameters in the scalar potential to have 
CP-violating phases, the mass matrix for the neutral scalars assumes the 
general form
\begin{equation}
{\cal L}_{\rm mass} = 
\left( \begin{array}{cc|cc} \eta_1^0 & \eta_2^0 & \phi_1^0 & \phi_2^0 \end{array} \right) 
\left( \begin{array}{c|c} {\cal M}_P^2  &  {\cal M}_{SP}^2 \\  \\  \hline & \\ 
             \left[{\cal M}_{SP}^2\right]^T & {\cal M}_S^2 \end{array} 
\right) \left( \begin{array}{c} \eta_1^0 \\ \eta_2^0 \\ \hline \phi_1^0 
\\ \phi_2^0 \end{array} \right) 
\end{equation} 
This $4 \times 4$ mass matrix is partitioned into $2 \times 2$ blocks, 
with independent ${\cal M}_P^2$, ${\cal M}_S^2$ and ${\cal M}_{SP}^2$ 
--- of which the last is absent in the CP-conserving MSSM but is 
generated in the CP-violating MSSM through the one-loop corrections 
mentioned above\cite{CPv1,CPv2,CPv3,CPv4,CPv5,CPv6}. The magnitude of 
different contributions to the terms in the $2 \times 2$ matrix ${\cal 
M}_{SP}^2$ may be estimated as \cite{CPv2}:
\beq
{\cal M}^2_{\rm SP} \approx {\cal O}\left ( \frac{M^4_t \mid \mu \mid 
\mid A_t \mid}{v^2 32 \pi^2 M^2_{\rm SUSY}}\right ) \sin \Phi_{\rm CP} 
\times \left [6, \frac{\mid A_t \mid^2 }{M^2_{\rm SUSY}}, \frac{\mid 
\mu\mid^2}{\tan\beta M^2_{\rm SUSY}}, \frac{\sin 2\Phi_{\rm CP}\mid 
A_t\mid \mid\mu\mid }{\sin \Phi_{\rm CP} M^2_{\rm SUSY}}\right ]
\enq
where $\Phi_{\rm CP} = {\rm Arg}(A_t\mu)$, $v = 246$ GeV. and the mass 
scale $M_{\rm SUSY}$ is defined by
\begin{equation}
M_{\rm SUSY}^2 = \frac{m^2_{\tilde t_1} + m^2_{\tilde t_2}}{2} \ .
\end{equation}
Rough estimates of the degree of CP violation in the Higgs sector can be 
formed by taking the dominant one(s) of these contributions. For 
example, from the above expression it is clear that a sizeable 
scalar-pseudoscalar mixing is possible for a large CP-violating phase 
$\Phi_{\rm CP}$, $|\mu|$ and $|A_t| > M_{\rm SUSY}$.

The diagonalisation of this $4 \times 4$ mass-squared matrix can be 
carried out in two stages, of which the first is to simply diagonalise 
the sub-matrix ${\cal M}_P^2$ and replace the pseudoscalars 
$\eta_{1}^0$, $\eta_2^0$ with the more familiar $G^0, A^0$.  It turns 
out that after this is done, the first row and column of the $4 \times 
4$ mass matrix are left only with contributions from tadpole diagrams, 
which would be removed in any renormalisation programme. It follows 
that, apart from a massless Goldsone boson $G^0$ which does not mix 
further with the other neutral states and is eventually absorbed by the 
massive $Z^0$ boson, we obtain a $3 \times 3$ Higgs mass-squared matrix 
${\cal M}^2$, with a mass term
\begin{equation}   
{\cal L}_{\rm mass} = 
\left( \begin{array}{ccc} A^0 & \phi_1^0 & \phi_2^0 \end{array} \right) 
\left( \begin{array}{ccc}        &                  &                \\
                                                   &  {\cal M}^2_{ij}   &           \\
                                                   &                  &                \\ \end{array} \right)
\left( \begin{array}{c} A^0 \\ \phi_1^0 \\ \phi_2^0 \end{array} \right) 
\end{equation}  
where $A^0$ is the appropriate eigenstate of ${\cal M}_P^2$. 
Diagonalising this $3\times 3$ symmetric matrix ${\cal M}^2_{ij}$ by an 
orthogonal matrix ${\cal O}$,
\begin{equation}
\left( \begin{array}{c} h^0_3 \\ h^0_2 \\ h_1^0 \end{array} \right)  = 
\left( \begin{array}{ccc}        &                  &                \\
                                                   &  {\cal O}_{ij}   &           \\
                                                   &                  &                \\\end{array} \right)
\left( \begin{array}{c} A^0 \\ \phi_1^0 \\ \phi_2^0 \end{array} \right) 
\end{equation}
we see that the physical mass eigenstates $h^0_1, h^0_2 $ and $h^0_3$ 
(in ascending order of mass) are mixtures of the CP-odd $A^0$ and the 
CP-even $\phi_1^0$ and $\phi_2^0$. These, therefore, are states of 
indefinite CP.  The usual sum rules for neutral Higgs boson masses (i.e. 
eigenvalues of ${\cal M}^2$) become much more complicated than in the 
CP-conserving case. Moreover, as $A^0$ is no longer a physical state, 
the charged Higgs boson mass $M_{H^\pm}$ is a more appropriate parameter 
for description of the MSSM Higgs-sector in place of the $M_A$ used in 
the CP-conserving model.

The coupling of these new states $h^0_i ~(i = 1,2,3)$ to the weak gauge 
bosons $W^\pm$ and $Z^0$ will obviously be different from those in the 
CP-conserving MSSM. We can write them as \cite{CPv2}
\begin{eqnarray}
{\cal L}_{hVV}^{\rm int} & = & g M_W \sum_{i=1}^3 g_{h_i VV} \left( h^0_iW_\mu^+W^{-,\mu} 
+\frac{1}{2 \cos^2\theta_W} h^0_iZ_\mu Z^{\mu} \right) 
\nonumber \\
{\cal L}_{hhZ}^{\rm int} & = & \frac{g}{2 \cos\theta_W} \sum_{i,j=1}^3 g_{h_ih_jZ} ( h^0_i\, \!\! \stackrel{\leftrightarrow}{\vspace{2pt}\partial}_{\!\mu} h^0_j )Z^\mu     + {\rm H.c.} \hspace{0.65in}
\nonumber \\
{\cal L}_{hH^\mp W^\pm}^{\rm int}& = & \frac{g}{2 \cos\theta_W} \sum_{i=1}^3 g_{h_iH^+W^-} ( h^0_i\, \!\!
\stackrel{\leftrightarrow}{\vspace{2pt}\partial}_{\!\mu} H^+)W^{-,\mu}   + {\rm H.c.}
\end{eqnarray}
where
\vspace*{-0.2in}
\begin{eqnarray}
g_{h_iVV} & = & {\cal O}_{1i}\cos\beta+ {\cal O}_{2i}\sin\beta
\nonumber \\
g_{h_ih_j Z} & = & {\cal O}_{3i}(\cos\beta O_{2j}- \sin\beta {\cal O}_{1j}) -(i\leftrightarrow j)
\nonumber \\
g_{h_iH^+W^-} & = & {\cal O}_{2i}\cos\beta- O_{1i}\sin\beta + i {\cal O}_{3i}
\end{eqnarray}
These couplings obey the following sum rules:
\begin{equation}
\sum_{i=1}^3 g^2_{h_i VV} = 1, \qquad\qquad
g^2_{h_i VV} + \mid g_{h_i H^+W^-}\mid^2 = 1, \qquad\qquad
g_{h_k VV} = \epsilon_{ijk} g_{h_i h_j Z} 
\label{eqn:sumrules}
\end{equation}
from which one can see that if two of the $g_{h^0_i ZZ}$ are known, then 
the whole set of couplings of the neutral Higgs boson to the gauge 
bosons are determined. It is interesting to see from 
Eqn.~(\ref{eqn:sumrules}) that in the case of large scalar-pseudoscalar 
mixing the suppressed $h^0_1 VV$ coupling means an enhanced $h^0_1 
H^+W^-$ coupling. This has been exploited to develop search strategies 
in Ref.~\cite{Ghosh:2004cc, Ghosh:2004wr}.

Having described the formalism in which one can study the mixed CP Higgs 
bosons of the model, we require to choose the parameters of interest. We 
use the package {\sc CPsuperH}(version 2.2) \cite{Lee:2007gn} to 
generate the entire particle spectrum of the CP-violating MSSM for every 
given set of input parameters. It has already been mentioned that the 
quantity $\sin \Phi_{\rm CP}/M^2_{\rm SUSY}$ needs to be large to 
support significant CP-mixing in the Higgs sector. As mentioned in the 
introductory section, the benchmark scenario dubbed the `CPX 
scenario'~\cite{CPv5} nicely showcases this CP violation since, among 
other things, it makes the $h^0_1ZZ$ coupling small enough to evade the 
LEP bounds~\cite{carena,Abbiendi:2004ww,lep2mssm}. This CPX scenario may 
be summarised as the specific parameter choices
\begin{eqnarray}
M_{\tilde{Q}} = M_{\tilde{t}} = M_{\tilde{b}} 
= \frac{\mu}{4} = \frac{|A_t|}{2} = \frac{| A_b|}{2} = \frac{| A_\tau|}{2}
&=&M_{\rm SUSY} \nonumber \\  
{\rm Arg}[A_t] = {\rm Arg}[A_b] = {\rm Arg}[A_\tau] &&
\label{eqn:CPX}
\end{eqnarray} 
It is important to note that this is just a benchmark scenario which 
allows us to find parameter choices which lie in the `blind spot' or 
inaccessible region of previous collider experiments, and does not 
exhaustively cover the whole of the inaccessible parameter space. 
However, it suffices to provide a framework for predicting the existence 
of light, hitherto-invisible Higgs bosons, which is the subject of this 
work. In fact, instead of exploring the entire CPX scenario and its 
manifold variations, we find that it is sufficient to choose four 
benchmark points BP-1, BP-2, BP-3 and BP-4, and concentrate our 
exploratory studies on these points. We use the well-known code {\sc 
HiggsBounds} (version 2.1.1) \cite{Bechtle:2008jh} to ensure that these 
four benchmark points are allowed by the LEP-2 and Tevatron Higgs 
searches. The choice of parameters common to all the four benchmark 
points (BP) are as follows: \vspace*{-0.2in} \begin{itemize} \item 
$M_{\rm SUSY} = 500$~GeV and $\Phi_{\rm CP} = {\rm Arg}[A_t] = {\rm 
Arg}[M_{\tilde g}]= \pi/2$. The related parameters are then fixed using 
Eqn.~(\ref{eqn:CPX}). \item The remaining phases are fixed as ${\rm 
Arg}[M_1] = {\rm Arg}[M_2] = 0$. \item The gluino mass is fixed to 
$M_{\tilde g} = 1.2$~TeV, as are the masses of squarks of the first two 
generations; the squarks of the third generation are assumed to have 
masses 500~GeV. \item The top quark mass is taken to be 
173.13~GeV\cite{CDFtop}. Of course, this is not a free parameter, but 
its exact value determines the top Yukawa coupling and hence controls 
the running of SUSY masses and couplings between the SUSY-breaking scale 
and the electroweak scale in a critical manner. \end{itemize} 
\vspace*{-0.2in} The parameters which differ from one benchmark point to 
another are given in Table~\ref{tab:BP1234}, where a vertical line 
separates these input parameters from some masses calculated by the {\sc 
CPsuperH} package.

\footnotesize
\begin{table}[htb]
\centering 
\begin{tabular}{c|cccc|ccc|rc}
\hline
BP & $M_1$ & $M_2$ & $\tan\beta$ & $M(H^\pm)$ & $M(h^0_1)$ & $M(h^0_2)$ 
& $M(h^0_3)$ & $M(\tilde{\chi}_1^0)$ & $M(\tilde{\chi}_1^\pm)$ \\
\hline\hline
1 & 100 & 200 &   6 & 125.7 & 49.4 & 101.8 & 130.4 &   99.7 & 198.6\\ [-2mm]
2 & 100 & 200 & 15 & 128.0 & 68.3 & 111.6 & 125.1 &   99.8 & 199.2 \\ [-2mm]
3 & 200 & 400 & 15 & 130.0 & 72.1 & 113.2 & 125.8 & 199.8 & 398.9 \\ [-2mm]
4 & 100 & 200 &   8 & 140.0 & 83.0 & 112.7 & 135.6 &   99.7 & 198.9 \\
\hline
\end{tabular}
\def\baselinestretch{1.1}
\caption{\footnotesize Our choice of benchmark points in the CPX 
scenario. In addition to the parameter choices, we present the masses of 
the three neutral Higgs states $h^0_1$, $h^0_2$ and $h^0_3$, as well as 
the lightest neutralino $\tilde{\chi}_1^0$ and the lightest chargino 
$\tilde{\chi}_1^\pm$. The next-to-lightest neutralino $\tilde{\chi}_2^0$ 
is almost degenerate with the $\tilde{\chi}_1^\pm$. All masses are in 
units of GeV.}
\def\baselinestretch{1.0}
\label{tab:BP1234}
\end{table}
\normalsize
This table illustrates very well some of the points made in the previous 
discussion. They have the common feature that the lightest Higgs boson 
is much lighter than the LEP bound and will therefore be highly boosted 
at the LHC. It is interesting that for BP-2 and BP-3, the third of the 
neutral Higgs bosons lies precisely in the range where the new boson has 
been found \cite{Higgs_discovery}, but it would be premature to read too 
much into this until we have a precise measurement of the couplings. 
Similarly we do not concern ourselves overly with the relatively light 
charged Higgs bosons, although they do make substantial contributions 
\cite{Lee} to the decay width for the process $B_s \to \mu^+\mu^-$. In 
fact, strictly speaking, if we take the current upper bounds from the 
LHCb on this branching ratio \cite{LHCb}, none of our benchmark points 
would be tenable\footnote{However, this can be circumvented in various 
ways, either by going beyond the minimal flavour violation paradigm, or 
by activating some of the supersymmetric phases set to zero in our 
choice of benchmark points, without prejudice to the existence of light 
neutral Higgs states. This issue will be taken up in a future work 
\cite{selves}.}. For the moment, however, our interest focuses on the 
light Higgs bosons $h^0_1$ (and sometimes the $h^0_2$), where we shall 
presently show that tagging of their boosted decay products can help 
their detection at the LHC, and thereby shed some light on the parameter 
space which has hitherto been a `blind spot' for collider experiments.

\vspace*{-0.2in}
\section{Boosted Higgs Bosons and Jet Substructure}

\vspace*{-0.2in} In this section we discuss the technique used to 
identify light Higgs bosons decaying into a pair of $b$-jets over and 
above the enormous QCD background. This is a technique based essentially 
on kinematics and hence is not specific to any model. The $CP$-violating 
MSSM described in the previous section acts here only as a 
phenomenologically-viable framework which can support the existence of 
light Higgs bosons.

The study of jet substructure began some years ago \cite{Seymour, 
Butterworth:2007ke, BDRS,Thaler:2008ju,Kaplan:2008ie,rev} with the 
realisation that in searches for new physics, the kinematic 
configuration of final states involving hadronic jets at the LHC can be 
very different from those studied earlier at colliders operating close 
to the electroweak scale, such as the LEP (91 -- 205~GeV), the Tevatron 
(around 300 -- 350~GeV) and the HERA (200 -- 300~GeV). At these 
machines, the major searches were for new particles of mass between some 
tens of GeV to a few hundred GeV, i.e. the masses were comparable to the 
machine energy. When particles of such mass are produced, they do not 
carry much momentum and hence are only mildly boosted. At the LHC, 
however, the available centre-of-mass energy is around 1 -- 2~TeV, but 
the particles being sought are the same as before. It follows, 
therefore, that, if produced, these particles will be very highly 
boosted. This realisation has led to a new paradigm for studies of new 
physics in hadronic final states at the LHC. In fact, if a new particle 
of mass in the range of a few hundred GeV is discovered at the LHC then 
further studies of that particle would almost always include a strong boost. It 
may be kept in mind, however, that if new physics continues to elude LHC 
searches, the mass limits on these particles will eventually increase 
and become comparable to the available machine energy --- thereby 
restoring the situation at earlier colliders and reinstating the 
techniques invented for those specific studies. The present study (and 
all similar studies) are, therefore, currently relevant because we are 
in the early stages of the LHC run.

The main feature of the decay products of a boosted particle decaying 
into multiple hadronic jets is that the final states remain highly 
collimated, appearing as a single fat jet. Thus, at the LHC, there will 
be $W/Z$-jets \cite{Seymour,Butterworth:2007ke,wz}, $t$-jets 
\cite{Thaler:2008ju,Kaplan:2008ie,top} and $H$-jets \cite{BDRS, higgs} 
in the Standard Model, and in new physics models there will be objects 
like charged Higgs jets \cite{chargedh}, neutralino jets 
\cite{neutralino}, etc., depending on the model chosen. The invariant 
mass of such a jet, constructed by adding the momenta of all the 
hadronic clusters, will peak at the mass of the parent particle, i.e. at 
$M_W$, $M_Z$, $m_t$, and so on. However, it is not enough to identify 
fat jets with a certain range of invariant mass, since there will be a 
substantial QCD background to these -- enough to mask the small numbers 
due to production of these heavier particles. It is necessary, 
therefore, to tag the fat jets further by scanning them at higher 
angular resolutions for substructures which would betray their origin 
from the decay of heavy particles, rather than a series of gluon 
radiations and gluon splittings, which characterises the typical QCD 
jet. Criteria imposed on the substructure of fat jets have been used 
with success to tag $W$, $Z$ and $t$ jets, and here we apply a set of 
such criteria designed to identify fat jets arising from light Higgs 
bosons decaying to a pair of $b$ quarks.

The exact method used by us to tag Higgs boson jets closely follows that 
used in Ref.~\cite{BDRS}. We now describe this technique, which has 
three stages. In our numerical simulation the basic processes are 
calculated using the well-known Monte Carlo generator {\sc Pythia} 
\cite{pythia}, and the jets are identified using the add-on package {\sc 
FastJet} \cite{fastjet}. Thus, we start with a bunch of hadronic final 
states, of which some are identifiable with known hadrons, but others 
are just clumps of quarks and gluons, bound together for some short 
lifetimes. This is the theoretical equivalent of what the 
experimentalist would see as a bunch of hadronic `clusters' recorded in 
the hadron calorimeter (HCAL).  As the first stage in our analysis, we 
identify a jet from these putative `clusters', by applying the 
Cambridge/Aachen (C/A) algorithm \cite{ca}, which operates as follows.
\vspace*{-0.2in}
\begin{itemize}
\item The angular distance $\Delta R_{ij}$ between all pairs ($i,j$) of 
`clusters' is given by
\begin{equation}
\Delta R_{ij} = \sqrt{(y_i - y_j)^2 +  (\varphi_i - \varphi_j)^2}
\label{eqn:CAdistance}
\end{equation}
where $y_i = \frac{1}{2}\ln(E_i-p_{Li})/(E_i+p_{Li})$ is the rapidity 
and $\varphi_i$ is the azimuthal angle of the $i^{\rm th}$ cluster. This 
quantity, which is invariant under longitudinal boosts, is tabulated for 
all pairs $(i,j)$ and the pair with the smallest value of $\Delta 
R_{ij}$ is merged into a single `cluster', i.e. the momentum four 
vectors are added into a single momentum four vector. We thus have a new 
configuration with one less `cluster' than before. \item The above 
exercise is then repeated with the new configuration, as a result of 
which the number of `clusters' is again reduced by unity. This is 
iterated until there remains only a single four vector which has been 
built up by this merging process, with the nearest hadronic `cluster' 
lying at an angular distance $\Delta R > R_0$, where $R_0$ is a 
predetermined angular width for a `fat' jet. The synthetic four vector 
($P,M$) built up by this process is identified as that of a jet, and the 
three-momentum $\vec{P}$ of this jet is identified as the jet axis. The 
invariant mass of this jet is simply constructed by taking $M = 
\sqrt{P_0^2 - \vec{P}^2}$. For Higgs boson masses in the range of a few 
tens of GeV the entire Higgs jet algorithm works best if we set $R_0 = 
0.6$ and this is what is done in the rest of the discussion. \item One 
then iterates the above procedure for all hadronic clusters lying 
outside the cone $\Delta R = R_0$ centred around this jet axis, i.e. for 
this part of the study, the four-vector already identified as a jet will 
be excluded.  It is likely that the remaining clusters will again 
combine to form a different jet under this algorithm, or more than one 
jet, as the iterative process continues. \item Some of the jets thus 
synthesised may be soft, with $P_T$ lying below 20~GeV. These are 
usually discounted and only hard jets ($P_T > 20$~GeV) thrown up by the 
C/A algorithm are involved in the remaining part of the analysis. 
\end{itemize}

\vspace*{-0.2in} The C/A algorithm described above\footnote{Many 
experimental analyses prefer to apply the so-called $k_T$ and anti-$k_T$ 
algorithms, where the C/A distance function $\Delta R_{ij}$ is modified by a 
momentum-dependent factor. For the present analysis, the anti-$k_T$ 
algorithm would not work at all, and there is no practical advantage to 
be gained by the $k_T$ algorithm. Hence, we confine ourselves to the 
simpler C/A algorithm.} generically leads to a multi-jet final state 
depending on the number and configuration of hadronic `clusters'. Of 
course, this algorithm will indiscriminately pick up QCD jets as well as 
those arising from heavy particle decay, so it should be considered only 
as the initial step in our search for heavy particle jets. In order that 
the next step be possible, the merging history must be stored for every 
hard `fat' jet created by the C/A algorithm. In the second stage of our 
analysis, this history is then used to (partially) eliminate QCD jets by 
{\it reversing} the process of synthesis and applying the following 
criteria at the different stages of the reversal:

\vspace*{-0.2in}
\begin{itemize}
\item The jet --- read four-momentum ($P,M$) --- is broken into two 
sub-jets ($p,m$) and ($p',m'$) by reversing the last merging performed 
in the C/A synthesis, which means that $P = p + p'$. The ordering should 
be such that $M \geq m > m'$.
\item We then check the `mass drop' i.e. if $m' < x M$ (where $x$ is a 
smallish fraction), then it is likely that the corresponding softer 
sub-jet was a radiated gluon. Following Ref.~\cite{BDRS}, we set $x = 
\frac{1}{3}$. There are two cases, viz.
\begin{enumerate}
\item If $0 < m' < x M$, then we eliminate the final step in the merging 
and zero in on the harder sub-jet with ($p,m$) and start again, i.e. we 
consider the previous step in which ($p,m$) was formed by the merging of 
two `clusters', and repeat this analysis. For all 
practical purposes, the softer sub-jet with ($p',m'$) forms no part of 
the remaining analysis.
\item If $xM < m' < M$, then we assume that we have found two sub-jets 
of comparable momenta and further apply a symmetry criterion. This 
amounts to calculating $\Delta R$ between the two sub-jets and 
constructing
\begin{equation}
y = \frac{{\rm min}(p_T^2, {p'}_T^2)}{{m'}^2} \Delta R^2 
\approx \frac{{\rm min}(p_T^2, {p'}_T^2)}{{\rm max}(p_T^2, {p'}_T^2)}
\label{eqn:BDRSasymmetry}
\end{equation}
The symmetry criterion is then set Ref.~\cite{BDRS} as $y > 0.09$, which 
is consistent with the value of $x$ chosen before. In practice, passing 
both these criteria simply means that we are satisfied that the sub-jets 
arise from decay of a heavy particle (as opposed to a gluon 
radiation/splitting) if they share the energy and momentum of the 
composite `fat' jet in a ratio not more skewed than 2:1.
\end{enumerate} 
\item The final step in identifying a `Higgs jet' would be to demand 
that the two reasonably symmetric sub-jets picked out by the above 
algorithm are tagged as $b$-jets. For the experimentalist, this would 
mean that these jets correspond to displaced vertices in the tracker. 
However, in a theoretical analysis, we have the luxury of knowing the 
exact components of each sub-jet, whose history can be tracked down to 
the parton level even before fragmentation. Thus, we can achieve virtual 
$b$-tagging simply by looking for a $b$-quark parentage for the sub-jets 
in question. This process, however, has a 100\% efficiency and a very 
small mistagging fraction, both of which are unrealistic. We therefore 
require to import proper efficiency criteria from the existing 
experimental analyses (see below).
\end{itemize} 
\vspace*{-0.2in} The above criteria are fairly efficient in eliminating 
QCD jets, but are not so efficient in eliminating jets arising from 
underlying events -- which are again a new problem at the enhanced 
energies and luminosities of the LHC. If these are not filtered out, the 
Higgs boson mass reconstruction is adversely affected. We note that the 
$b$~sub-jets will have an angular separation given approximately by 
$R_{b\bar{b}} \gsim 2M_H/p_T^H$ -- which is smallish for boosted Higgs 
bosons, but much larger than the angular resolution of the ATLAS and CMS 
detectors. Although the cross section for $b\bar{b}$ pairs from 
underlying events will scale as $(R_{b\bar{b}})^4$ \cite{r4}, it turns 
out that for Higgs masses around 100~GeV and $p_T \sim 200 - 300$~GeV, 
$R_{b\bar{b}}$ is not so small that underlying events can be neglected. 
Therefore, there is a third stage to our analysis, where the Higgs jet 
candidates picked out by the previous algorithm are subjected to a 
`filtering' process. This works as follows.
\vspace*{-0.2in}
\begin{itemize}
\item Working back as before, we decompose the constituents of the 
candidate jet into the original hadronic `clusters'. Other jets and 
stable final states remain untouched.
\item We define a new angular criterion $r_0$, and re-run the C/A 
algorithm to identify sub-jets with this criterion instead of $R_0$. 
Again following Ref.~\cite{BDRS}, we set $r_0 = 0.2$. However, as a 
check, we tried different values of this cone radius, viz. $r_0 = 0.1, 
0.2, 0.3$ etc. For Higgs bosons in the mass range of interest, the 
maximum efficiency in choosing candidates for light Higgs bosons 
occurs for the choice $r_0 = 0.2$, and this is what we present our 
results for.
\item Of all the sub-jets thrown up by the above algorithm we choose the 
three hardest and eliminate all the others. Of these three sub-jets, if 
the hardest two are $b$-tagged, we identify the jet as a Higgs boson
candidate. Note that we take three hard sub-jets in order not to lose the 
not-so-rare $H \to b\bar{b}g$ events.
\end{itemize}
\vspace*{-0.2in} 
The above filtering process removes rare QCD cases where a very hard 
gluon is radiated, as well as hard emissions from underlying events, 
which could have satisfied the criteria in the first two stages, but 
will fail the last one. Once these three stages of identification are 
passed, we can be sure that a reasonable fraction of the tagged jets 
will originate from Higgs bosons. Our next step, therefore, is to see 
how efficient this Higgs jet-finding algorithm can be.

\begin{center}
\begin{figure}[htb] 
\includegraphics[height=0.44\textheight,width=1.1\textwidth]{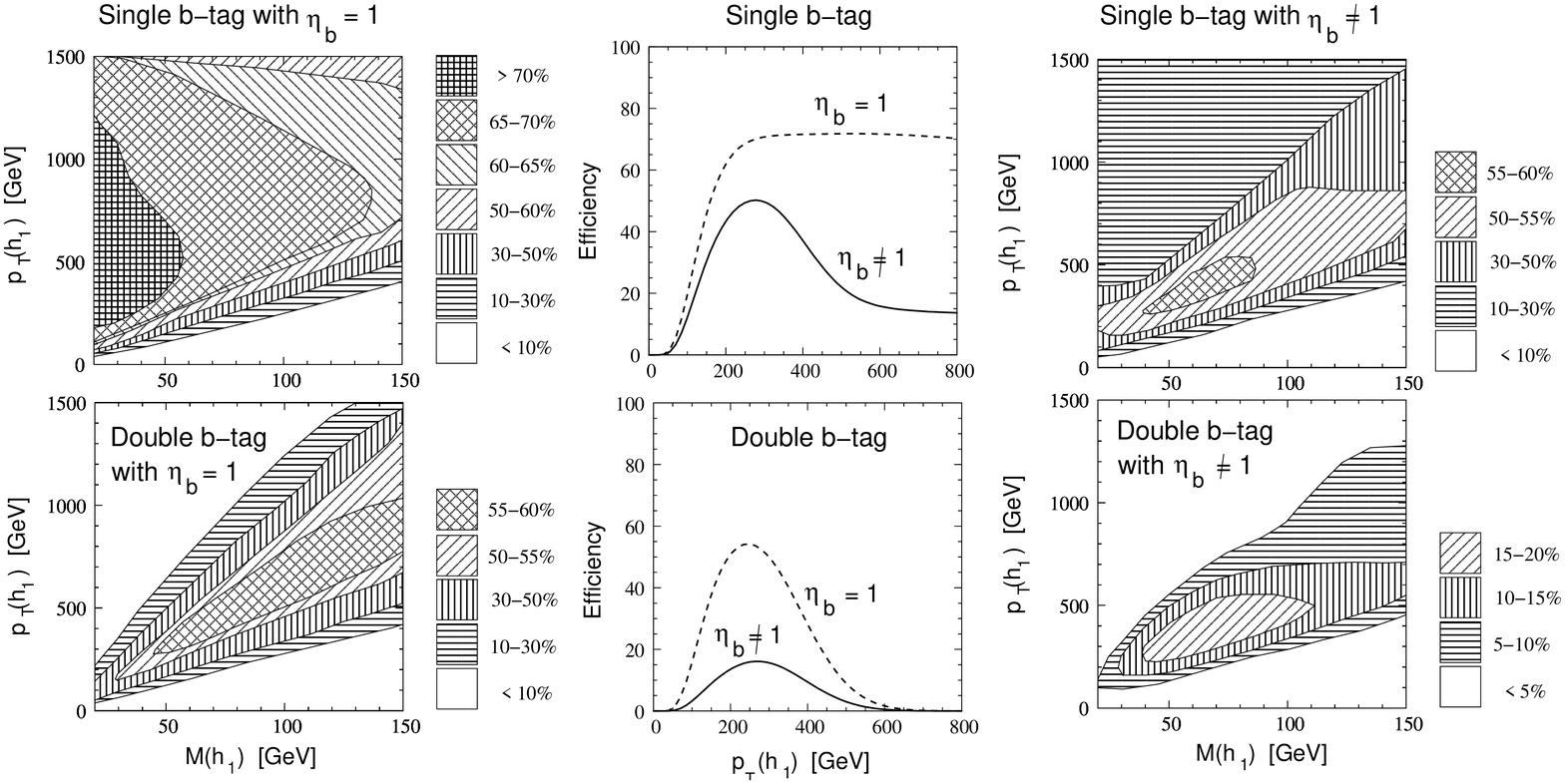} 
\vspace*{-0.2in}
\def\baselinestretch{1.1}
\caption{\footnotesize Illustrating the efficiency of Higgs-tagging of 
`fat' jets by the method described in this section. The upper (lower) 
row of panels corresponds to a single (double) $b$-tag among the 
sub-jets thrown up by the `filtering' algorithm. The two panels on the 
left illustrate contours of the efficiency of this algorithm in the 
plane of $p_T(h_1)$ versus $M(h_1)$ assuming $b$-tagging is perfect. 
The panels in the middle show the effect, for $M(h_1) = 100$~GeV, of 
including the $b$-tagging efficiency with its $p_T$-dependence. The 
panels on the right show the results of convolution of the $b$-tagging 
efficiencies with those shown in the panels on the left, thereby 
producing a more-or-less realistic estimate of the actual efficiencies 
obtainable at the LHC.  }
\def\baselinestretch{1.25}
\label{fig:efficiency}
\end{figure}
\end{center}

To make a specific study, we have made a Monte Carlo simulation of the 
following exclusive process:
\vspace*{-0.2in}
\begin{eqnarray}
p + p & \to & \tilde{\chi}_2^0 + \tilde{\chi}^0_1 \nonumber \\
         &       &  \hookrightarrow \tilde{\chi}^0_1 + h_1^0
\label{eqn:toymodel}
\end{eqnarray}
where the branching ratio for $\tilde{\chi}_2^0 \to \tilde{\chi}^0_1 + 
h_1^0$ is set to unity. The masses of the neutralinos $\tilde{\chi}^0_1$ 
and $\tilde{\chi}^0_2$ are set to 100~GeV and 700~GeV respectively in 
this toy model, and all other processes are switched off. The masses of 
the $\tilde{\chi}^0_1$ and $\tilde{\chi}^0_2$ are chosen so as to have a 
sufficient phase space for the Higgs boson to be produced with a large 
$p_T$. It may be noted that it was not essential to have a 
supersymmetric origin of the Higgs bosons -- a Higgstrahlung process $p 
+ p \to Z^0 + h^0_1$, with subsequent decay of $Z^0 \to \nu \bar{\nu}$, 
would have yielded the same final state, where we have only $h_1^0$ 
particles decaying to $b\bar{b}$ pairs in addition to other incidental 
debris from effects like initial state radiation, final state radiation, 
multiple interactions, and, of course, a great deal of missing 
energy and momentum. However, the processes in Eqn.~(\ref{eqn:toymodel}) 
are preferred since they permit us to study a larger range in $p_T$ for 
the Higgs boson than the Higgstrahlung process would have allowed. In 
this simulation, the mass of the Higgs boson $h_1^0$ is varied over the 
range 20 -- 150~GeV and the $p_T$ is noted. A total of 50\,000 `events' 
was simulated for every value of $M(h_1)$. For every bin in $p_T(h_1)$, 
we apply the jet-finding algorithm described above in all its three 
stages and note the efficiency, i.e the ratio of the number actually 
tagged as Higgs jets to the original number of $h_1^0$ produced. This 
enables us to produce the contour plots of Figure~\ref{fig:efficiency}.

\vspace{-0.2in}

\vspace{-0.7in}
Some explanation of the plots in Figure~\ref{fig:efficiency} is required 
at this stage. Let us first consider the upper row of panels in 
Figure~\ref{fig:efficiency}. We start with the upper panel on the 
extreme left, labelled `Single $b$-tag with $\eta_b = 1$.' This shows 
the result of implementation of the above algorithm with just one 
difference: after `filtering', we pick up three hard jets and of the 
hardest two, only one is required to have a $b$-tag. The plot shows 
contours of efficiency for the Higgs-jet finding algorithm, with the 
$b$-tagging efficiency set to unity. It may be seen that for a given 
Higgs mass $M(h_1)$, the efficiency is small for small $p_T$, increases 
as $p_T$ increases, and then falls again as $p_T$ increases to very 
large values. This behaviour is exactly what one would expect as the 
boost parameter increases. When the boost is very small, the jets 
arising from Higgs boson decay tend to be widely separated, and hence 
jets found within a cone of radius $R_0 = 0.6$ will not show any 
substructure, as they essentially arise from the decay of a single 
$b$-quark. As the boost increases, more of the jets will now get 
collimated and we will begin to see the phenomenon of a fat jet with 
substructure, to which the finding algorithm is tuned. Not surprisingly, 
the efficiency will increase in this regime. However, when the boost is 
very large, the final states will form a very thin pencil and it may not 
be possible to resolve individual sub-jets using the criterion $r_0 = 
0.2$. In this case, the efficiency will again fall. Increase in the 
Higgs mass $M(h_1)$ is equivalent to scaling down the $p_T$, and this is 
reflected in the contours shown in the figure. It augurs well for the 
success of this method that there is a substantial range of Higgs mass 
and Higgs $p_T$ where the efficiency can be as high as 70\% or above, 
and that much of the $p_T(h_1)$--$M(h_1)$ plane corresponds to an 
efficiency of 50\% or more.

As mentioned before, it is unrealistic to take the 
$b$-tagging efficiency as $\eta_b = 1$. Following the experimental 
collaborations \cite{CMSbtag}, we choose the $p_T$-dependent efficiency 
as
$$
\eta_b = 
\left\{ \begin{array}{ll} 0~~ & {\rm for} \ \ p_T^b \leq 20~{\rm GeV} \\
                          0.3 & {\rm for} \ \ 20~{\rm GeV} < p_T^b \leq 50~{\rm GeV} \\
                          0.6 & {\rm for} \ \ 50~{\rm GeV} < p_T^b \leq 400~{\rm GeV} \\
                          0.2 & {\rm for} \ \ p_T > 400~{\rm GeV}
               \end{array} \right.   
$$ 
This must be convoluted with the non-realistic efficiency shown in the 
upper left panel of Figure~\ref{fig:efficiency}, to obtain a more 
realistic picture. The results of such a convolution are shown in the 
central upper panel, marked `Single $b$-tag' for $M(h_1) = 100$~GeV. 
Here, it is easy to see how the efficiency is reduced quite dramatically 
by the requirement of $b$-tagging. What we gain in return, of course, is 
a much sharper drop in the background (not shown). The panel on the 
upper right of Figure~\ref{fig:efficiency}, which is marked `Single 
$b$-tag with $\eta_b \neq 1$', shows the contours of Higgs tagging 
efficiency after convolution with the $b$-tagging efficiency.  This is 
a much more realistic estimate of the efficiency, where one notes that 
the poorer results at high $p_T$ arise because of the weakness of the 
$b$-tagging algorithm in that regime. There is still, however, a 
substantial range of Higgs mass and $p_T$ where the efficiency is 30\% 
or more, and this is enough for the Higgs tagging algorithm to yield 
positive results, as we shall presently demonstrate.

Obviously, if we demand that {\it both} the harder sub-jets be 
$b$-tagged, we will get even lower efficiencies. The advantage, on the 
other hand will be to have almost negligible mistagging probabilities. 
The lower three panels in Figure~1 illustrate the efficiency obtained by 
tagging both $b$-sub-jets. On the left, we plot contours of efficiency 
for a `Double $b$-tag with $\eta_b = 1$', where efficiencies above 55\% 
may be achieved in the optimum range of Higgs mass and transverse 
momentum. The central panel, marked `Double $b$-tag' illustrates, like 
the panel above it, the effect of implementing realistic $b$-tagging on 
the efficiency contours for $M(h_1) = 100$~GeV. Not only does the 
overall efficiency fall below some 20\% in the best case, but there is a 
sharp falling-off at high $p_T$, as expected. The realistic contours are 
shown in the panel on the lower right, marked `Double $b$-tag with 
$\eta_b \neq 1$, and here we have a rather modest region even if we 
demand an efficiency of 10\%, with only a small region revealing an 
efficiency of more than 15\%.

In view of the efficiency plots presented in Figure~\ref{fig:efficiency} 
and discussed above, we suggest that a single $b$-tag may be a more 
efficient way of identifying Higgs jets than the double $b$-tag.  Though 
our analysis will take both the cases into account, it will be seen that 
the single $b$-tag is better for the earlier runs of the LHC, while the 
double $b$-tag will serve to clinch the discovery -- if made -- in the 
later higher luminosity runs of the same machine.

A final comment is in order before we end this section and go on to our 
specific results. The analysis in this section, though showcased using 
the supersymmetric process in Equation~(\ref{eqn:toymodel}), is really 
quite model-independent and the efficiency plots in 
Figure~\ref{fig:efficiency} are completely general, independent of the 
processes that produce the light Higgs boson $h_1^0$.  Our analysis can, 
therefore, be extended not only to light Higgs bosons, but to any object 
of comparable mass decaying to a pair of jets. This includes the $W^\pm$ 
and $Z^0$ bosons, identifiable by their known masses, but also many 
exotics in some of the wide range of models which predict new physics.

\vspace*{-0.2in}
\section{Bump Hunting in Boosted Jets : CPX Scenario}

\vspace*{-0.2in} Having set up the mechanism to identify boosted Higgs 
jets, we now apply this to the model of interest, viz., the CP-violating 
MSSM. There are several mechanisms by which the Higgs bosons $h^0_i$ ($i 
= 1,2,3$) can be produced in this model. We list some of the 
parton-level processes below.
\vspace*{-0.2in}
\begin{itemize}
\item Gluon fusion: $g + g \to h_i^0$, which is the major production 
process in the SM. Here, however, the produced $h_i^0$ will have very 
small $p_T$.
\item Vector boson fusion: $V + V \to h_i^0$, where $V = W^\pm, Z^0$, 
another SM process, where the $h_i^0$ are produced in association with
two highly forward jets.
\item Higgstrahlung: $q + \bar{q}(\bar{q}') \to V^* \to V + h^0_i$, where 
$V = W^\pm, Z^0$, which is suppressed compared to the previous two, but 
produces $h_i^0$ with larger $p_T$.
\item Associated production with top quarks: $p + p \to t + \bar{t} + 
h^0_i$, which also has a small cross section, but produces $h^0_i$ with 
large $p_T$.
\item Associated production with gaugino states: $p + p \to 
\tilde{\chi}_{1,2}^\pm + \tilde{\chi}_{1,2}^\mp + h^0_i$ or 
$\tilde{\chi}_{1,2,3,4}^0 + \tilde{\chi}_{1,2,3,4}^0 + h^0_i$ or 
$\tilde{\chi}_{1,2}^\pm + \tilde{\chi}_{1,2,3,4}^0 + h^0_i$, which 
occur only in SUSY models.
\item Chargino decay: $\tilde{\chi}_2^\pm \to \tilde{\chi}_1^\pm 
+ h^0_i$ where the charginos are produced directly or in cascade decays of 
squarks/gluinos.
\item Neutralino decay: $\tilde{\chi}_{2,3,4}^\pm \to 
\tilde{\chi}_1^\pm + h^0_i$ where the neutralinos are produced directly 
or in cascade decays of squarks/gluinos.
\item Stop and sbottom decay: $\tilde{t}_2 \to \tilde{t}_1 + h_i^0$ and 
$\tilde{b}_2 \to \tilde{b}_1 + h_i^0$, where the heavier stop or sbottom 
can be either produced directly or arise as a product of gluino decay.
\item Associated production with stop states: $p + p \to \tilde{t}_{1,2} 
+ \tilde{t}_{1,2}^\ast + h^0_i$, which is similar to the associated 
production with top quarks.
\end{itemize}
\vspace*{-0.2in} The above list is illustrative, but not exhaustive, for 
there can be several stages in SUSY cascade decays where the $h^0_i$ can 
be produced. All these processes are, however, taken care of in our 
numerical simulation using {\sc Pythia}.

Obviously, with so many contributing processes the LHC will produce 
large numbers of the $h^0_i$ as the run progresses. The question which 
interests us, however, is whether these produced $h^0_i$'s will be 
detectable using the tagging algorithm described in the previous 
section. For this, we have seen that the crucial kinematic determinant 
is the transverse momentum $p_T$. It is important, therefore, to have a 
clear picture of the $p_T$ distribution of the three Higgs scalars 
$h^0_i$. Unfortunately, this distribution is very sensitive to the 
masses of the $h^0_i$ and the mass-splitting between the different 
chargino and neutralino states, i.e. on the exact choice of model 
parameters. Thus, it is very difficult to make very general statements 
about the nature of these $p_T$ distributions. Instead, we find it 
expedient to focus on the four benchmark points described in 
Table~\ref{tab:BP1234}. Since these are chosen to cover various aspects 
of the CP-violating MSSM parameter space and guarantee the existence of 
light Higgs states, we can expect the $p_T$ distributions of the $h^0_i$ 
at these points to carry information about the strengths and weaknesses 
of the Higgs-tagging algorithm.

\footnotesize
\begin{table}[htb]
\centering 
\begin{tabular}{l|cccc}
\hline
Process &  BP-1  & BP-2   &  BP-3   & BP-4 \\ \hline\hline
$\tilde{\chi}_2^0 \to \tilde{\chi}_1^0 + h_1^0$ &
99.6  & 98.8   & 10.4    & 98.4 \\ [-2mm]
$\tilde{\chi}_2^0 \to \tilde{\chi}_1^0 + h^0_2$ &
--  & --   & 71.2    & -- \\ [-2mm]
$\tilde{\chi}_2^0 \to \tilde{\chi}_1^0 + h^0_3$ &
--  & --   & 18.0    & -- \\ \hline
$h^0_3 \to h_1^0 + h_1^0$ & 84.1  & --   & --    & -- \\ [-2mm]
$h^0_2 \to h_1^0 + h_1^0$ & 51.9  & --   & --    & -- \\ \hline
$h^0_3 \to b + \bar{b}$   & 14.3 & 90.4 & 90.4 & 90.2 \\ [-2mm]
$h^0_2 \to b + \bar{b}$   & 43.7 & 90.7 & 90.7 & 90.5 \\ [-2mm]
$h^0_1 \to b + \bar{b}$   & 92.0 & 91.5 & 91.4 & 91.2 \\ \hline
$\tilde{t}_2 \to \tilde{t}_1 + h^0_1$ & 0.23 & 0.02 & 0.01 & 0.06 \\ [-2mm]
$\tilde{t}_2 \to \tilde{t}_1 + h^0_2$ & 29.2 & 31.2 & 26.9 & 14.7 \\ [-2mm]
$\tilde{t}_2 \to \tilde{t}_1 + h^0_3$ & 26.9 & 19.2 & 24.8 & 42.0 \\
\hline
\end{tabular}
\def\baselinestretch{1.1}
\caption{\footnotesize Some important branching ratios (per cent) for 
the four benchmark points of Table~\ref{tab:BP1234} in Section~2. Blank 
entries indicate that the corresponding decay is kinematically 
disallowed. Note that the $h_1^0$ always decay predominantly into 
$b\bar{b}$ pairs. }
\def\baselinestretch{1.25}
\label{tab:BR1234}
\end{table}
\normalsize

At the LHC, the production cross-section of the Higgs bosons depends not 
only on their masses, but also on their couplings and on the branching 
ratios of heavier particles to final states involving these Higgs 
particles. As there are many processes, this is not easy to predict or to 
explain. Some information about the principal production modes can be 
gleaned from a study of the branching ratios of gauginos to the Higgs 
states and the branching ratios of the latter to $b\bar{b}$ states. Some 
of these are listed in Table~\ref{tab:BR1234}. From the table, it is 
obvious that the $h^0_1$ can arise from the decay of heavy gaugino as 
well as Higgs states, and that it always decays dominantly to $b\bar{b}$ 
pairs -- as we have implicitly assumed in setting up the Higgs-tagging 
algorithm. The results shown in Table~\ref{tab:BR1234} also indicate that the 
benchmark point BP-3 will have qualitatively different features from the 
others, as we have already noted.

\vspace*{-0.2in}
\begin{center}
\begin{figure}[h] 
\center{\includegraphics[width=0.6\textwidth]{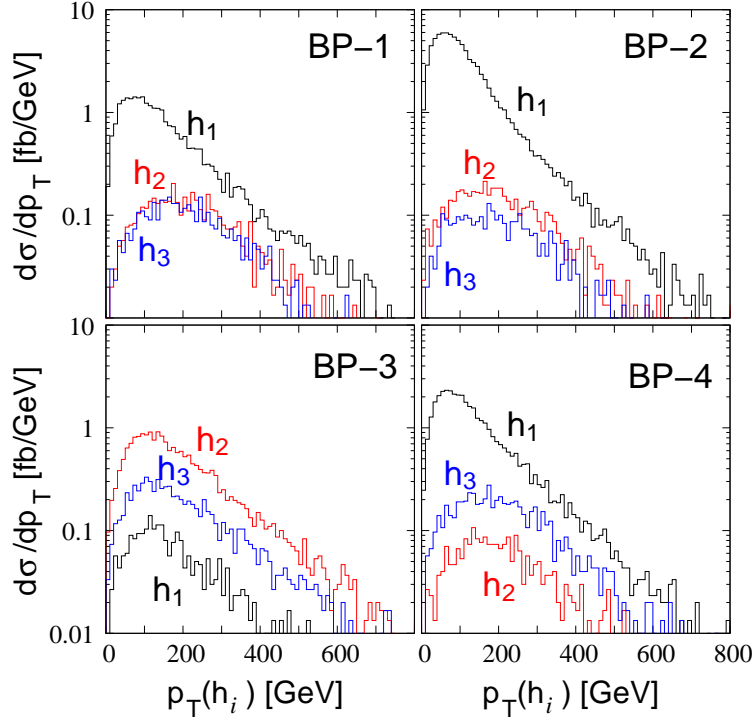}} 
\def\baselinestretch{1.1}
\caption{\footnotesize Distributions in transverse momentum $p_T(h_i)$ 
for the three Higgs bosons $h^0_i$ ($i = 1,2,3$) at the 14-TeV LHC.  
Black, red and blue histograms correspond, respectively, to the $h^0_1$, 
$h^0_2$ and $h^0_3$ states (as marked). The four panels in this plot are 
marked BP-1, BP-2, BP-3 and BP-4, corresponding to the parameter choices 
of Table~\ref{tab:BP1234}. No kinematic cuts were applied to generate 
these distributions.}
\def\baselinestretch{1.25}
\label{fig:pTdist1}
\end{figure}
\end{center}

\vspace*{-0.2in} In Figure~\ref{fig:pTdist1}, we generate the $p_T$ 
distributions for the three Higgs states. Black, red and blue histograms 
correspond, respectively, to the $h^0_1$, $h^0_2$ and $h^0_3$ states for 
each of the four benchmark points on the corresponding panel 
in Figure~\ref{fig:pTdist1}, as marked. Since these plots are generated for a 
theoretical understanding of the Higgs-tagging algorithm, none of the 
standard kinematic cuts used in SUSY searches have been applied. These 
plots immediately tell us that for BP-1, BP-2 and BP-4, the lightest 
$h^0_1$ is the most produced of the three states, whereas for the BP-3, 
it is the heavier $h^0_2$ which is produced most copiously.

We are now in a position to correlate the $p_T$ distributions of 
Figure~\ref{fig:pTdist1} with the efficiency plots of 
Figure~\ref{fig:efficiency}. Let us first consider the panel marked 
BP-1. In this case, the masses of the Higgs states are approximately 50, 
100 and 130 GeV respectively. If we consider the panel marked `Single 
$b$-tag with $\eta_b \neq 1$' in Figure~\ref{fig:efficiency}, we note 
that for the lightest state, we can obtain a tagging efficiency above 
30\% for the $p_T$ range $200 - 400$~GeV.  This is not the peak region 
of the $p_T$ distribution shown in Figure~\ref{fig:pTdist1}, but just to 
the right of it. It follows that the tagging algorithm will miss a 
fairly large fraction of the Higgs bosons, though it will still be able 
to capture a significant number. We shall see presently that this is 
enough to identify the Higgs state. For the moment, we focus on the 
other states, where we have to go to $p_T > 400$~GeV to obtain any 
useful efficiency. However, for such large values of $p_T$, the 
production cross section for these states is at least an order of 
magnitude smaller than that of the $h^0_1$ in the region where it can be 
tagged with reasonable efficiency. Thus, we should not expect any useful 
signal from the production of these states, i.e. they will remain 
invisible to our study. For such states, we will have to turn to the 
rare $\gamma\gamma$ decay mode, which is a much more difficult study and 
beyond the scope of this work.

For the other benchmark points, we can carry out a similar analysis 
based on a combination of Figures~\ref{fig:efficiency} and 
\ref{fig:pTdist1}. Despite the variation in parameters, the conclusions 
for BP-2 and BP-4 are very similar to that of BP-1, and hence are not 
detailed here. For BP-3, however, the situation is somewhat different. 
As in the case of BP-1, reasonable efficiencies for the heavy $h^0_2$ 
state ($~\sim 113$~GeV) take us into the high-$p_T$ regime, where we 
pick up only the tail of the $p_T$ distribution, whereas the lightest 
$h^0_1$ state ($~\sim 72$~GeV) can be tagged with better efficiency in 
the range where it is produced more copiously. However, the overall 
production of the $h^0_2$ is so much larger than that of the $h^0_1$ 
that we should expect the final cross section for the $h^0_2$ to be at 
least comparable with that of the $h^0_1$. We should not expect much 
from the $h^0_3$, which is neither light enough to avail of the higher 
efficiencies, nor is produced in enough numbers to offset that 
disadvantage.

If we now pass from the `Single $b$-tag with $\eta_b \neq 1$' to the 
`Double $b$-tag with $\eta_b \neq 1$', we shall obtain very similar 
results, except that the efficiencies will drop by a factor of 3 or 
more. Hence discovery through this mode will require the collection of 
more data than the previous case. However, as mentioned before, this 
should be treated as secondary, clinching evidence once we have a 
definite signal in the `Single $b$-tag with $\eta_b \neq 1$' data 
sample.

The results of Figures~\ref{fig:efficiency} and \ref{fig:pTdist1} and 
the arguments presented above make a reasonable case for using the 
tagging algorithm of Section~3 as a tool in a search for light Higgs 
bosons in the CP-violating MSSM. However, such a study now requires to 
be carried out with a full {\sc Pythia} simulation including all 
relevant processes and with jet identification through the {\sc FastJet} 
package, as explained in Section~3 for a toy model. We have carried out 
this study for the four benchmark points as above, after generating the 
SUSY mass spectrum and couplings using {\sc CPsuperH} as incorporated in 
the {\sc CalcHEP} package \cite{CalcHEP}. Parton density functions of 
the CTEQ5L set \cite{CTEQ5L} have been used and the factorisation scale 
is set to the parton-level energy. The final states of interest either 
have multijets and large missing $p_T$ ($\not{\!\!p}_T$) or one hard 
lepton, multijets and large $\not{\!\!p}_T$. After generating each 
event, we have applied the following kinematic cuts, which are 
more-or-less in line with those applied in standard SUSY searches by the 
CMS Collaboration \cite{CMSSusy}. These are listed below.

\vspace*{-0.2in}
\begin{enumerate}
\item We select events with jets ($J$) having transverse momentum $p_T^J
> 30$~GeV and pseudorapidity $|\eta_J| < 3$.  The identified jets will 
be labelled $J_1, J_2, \dots$ in order of decreasing $p_T^J$.
\item The events must have\footnote{Here we deviate a little from the 
CMS analysis, which has $\not{\!\!p}_T > 200$~GeV.} missing transverse 
momentum satisfying $\not{\!\!p}_T > 300$~GeV. To calculate this, 
we take into account all jets with $p_T^J > 20$~GeV and pseudorapidity 
$|\eta_J| < 4.5$ and all leptons with $p_T^\ell > 10$~GeV and 
pseudorapidity $|\eta_\ell| < 2.5$.
\item The leading (maximum $p_T$) jet must have pseudorapidity 
$|\eta_{J_1}| < 1.7$.
\item The two leading jets must have $p_T^{J_1} > 180$~GeV and 
$p_T^{J_2} > 110$~GeV.
\item We calculate an effective mass $M_{\rm eff} = p_T^{J_2} + 
p_T^{J_3} + p_T^{J_4} + \not{\!\!p}_T$ and impose the condition $M_{\rm 
eff} > 500$~GeV. Naturally, $p_T^{J_4}$ is added only if the 
corresponding jet exists.
\item The angular separation between the two leading jets and the 
$\not{\!\!p}_T$ are calculated as $R_1 = \sqrt{(\pi - \delta\varphi_1)^2 + 
\delta\varphi_2^2}$ and $R_2 = \sqrt{\delta\varphi_1^2 + (\pi - 
\delta\varphi_2)^2}$, where $\delta\varphi_1 = |\phi_{j_1} - 
\phi_{\not{\!\!p}_T}|$ and $\delta\varphi_2 = |\phi_{j_2} - 
\phi_{\not{\!\!\!p}_T}|$. On 
these, we impose the conditions that $R_{1,2} > 0.5$.
\item The azimuthal angular separation between all jets and the 
$\not{\!\!p}_T$ must satisfy the criterion $\delta\varphi_1 = |\phi_{j} 
- \phi_{\not{\!\!p}_T}| > 0.3$.
\item The azimuthal angular separation between the next-to-leading jet 
and the $\not{\!\!p}_T$ must satisfy the criterion $\delta\varphi_2 = 
|\phi_{j_2} - \phi_{\not{\!\!p}_T}| > 0.35$.
\item We impose a lepton veto, i.e. the event should not contain any 
isolated lepton with $p_T^\ell > 20$~GeV and $|\eta_\ell| < 2.5$. The 
isolation criteria imposed on each lepton are ($a$) that the angular 
separation $\Delta R_{\ell J}$ between the lepton and every jet should 
not be less than 0.4 and ($b$) that the sum of the scalar $p_T$ of all 
stable visible particles within a cone of radius $\Delta R = 0.2$ around 
the lepton should not exceed 10~GeV.
\end{enumerate}

\footnotesize
\begin{table}[htb]
\centering 
\begin{tabular}{c|rrrr|r}
\hline
        & BP-1       & BP-2        & BP-3       & BP-4       & $t\bar{t}$  \\ \hline\hline
 Sample & 114\,000   & 111\,090    & 108\,720   & 109\,170   & 11\,070\,000     \\ [-2mm]
Cut  1  &  103\,880  & 101\,968    &  80\,624   &  99\,910   &  7\,208\,031     \\ [-2mm]
Cut  2  &   11\,161  &  11\,539    &  13\,778   &  11\,012   &      19\,360    \\ [-2mm]
Cut  3  &   10\,489  &  10\,861    &  12\,726   &  10\,410   &      17\,156    \\ [-2mm]
Cut  4  &    9\,759  &  10\,041    &  10\,942   &   9\,695   &      12\,165    \\ [-2mm]
Cut  5  &    9\,745  &  10\,025    &  10\,896   &   9\,684   &      11\,997    \\ [-2mm]
Cut  6  &    8\,237  &   8\,517    &  9\,451    &   8\,225   &       7\,372    \\ [-2mm]
Cut  7  &    5\,220  &   5\,452    &  6\,335    &   5\,211   &       4\,697    \\ [-2mm]
Cut  8  &    5\,199  &   5\,431    &  6\,317    &   5\,193   &       4\,679    \\ [-2mm]
Cut  9  &    3\,436  &   3\,814    &  5\,089    &   3\,501   &       1\,880    \\ \hline
\end{tabular}
\def\baselinestretch{1.1}
\caption{\footnotesize Kinematic cut flow table for a luminosity of 
30~fb$^{-1}$ at each of the benchmark points BP-1 to BP-4. The cuts are 
numbered as in the text.  Note the drastic effect of Cut 2, which 
requires $\not{\!\!p}_T > 300$~GeV. The efficacy of these 
cuts in removing the enormous $t\bar{t}$ background is immediately 
obvious.}
\def\baselinestretch{1.25}
\label{tab:cutflow}
\end{table}
\normalsize

It is interesting to see how the raw signal plotted in 
Figure~\ref{fig:pTdist1} is affected by the different cuts enumerated 
above. The total SUSY cross sections are rather large, being at the 
level of 3.8~pb, 3.7~pb, 3.6~pb and 3.6~pb for the benchmark points 
BP-1, BP-2, BP-3 and BP-4 respectively, but are dwarfed by the 
$t\bar{t}$ cross section which is 369~pb at the lowest order. The 
effects of the different cuts on the cross section are displayed in 
Table~\ref{tab:cutflow}, where we assume an integrated luminosity of 
30~fb$^{-1}$. The numbering 1--9 of the cuts is exactly as in the text 
above. The crucial role of the Cut 2 (on $\not{\!\!p}_T$) is obvious, 
and may be taken as a justification of the choice of a stronger cut than 
that of the CMS collaboration. It is also interesting to note that it is 
the lepton veto which finally reduces the $t\bar{t}$ background to a 
tractable value without affecting the signal events quite as strongly.

\begin{center}
\begin{figure}[ht] 
\center{\includegraphics[height=0.57\textheight,width=0.65\textwidth]{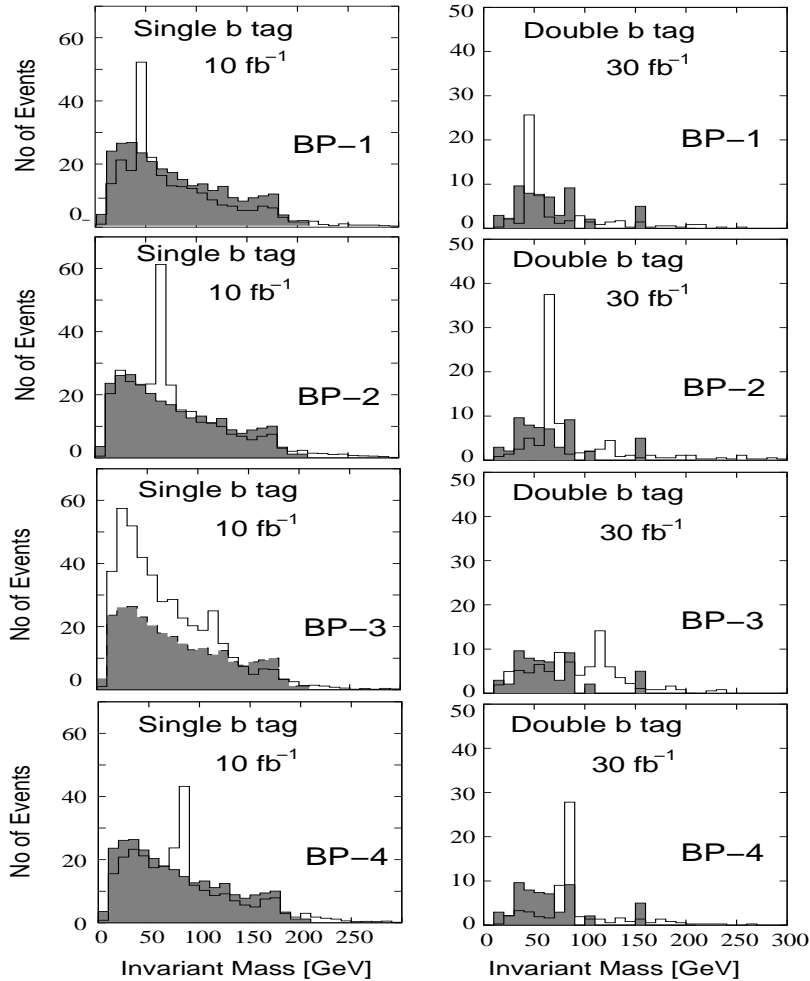}}
\def\baselinestretch{1.1}
\caption{\footnotesize Bin-wise invariant mass distribution of the 
leading jet with single (double) $b$-tags in 10 (30) fb$^{-1}$ of data 
at the 14 TeV LHC. The shaded histogram represents the $t\bar{t}$ 
background. Note the clear resonances corresponding to the lightest 
Higgs boson $h^0_1$ for the benchmark points BP-1, BP-2 and BP-4 (marked 
on the respective panels). For BP-3 and a single $b$-tag, we find a 
modest resonance corresponding to the $h^0_2$, while, for a double 
$b$-tag, there is only a tiny bump corresponding to the $h^0_2$. }
\def\baselinestretch{1.25}
\label{fig:CPX}
\end{figure}
\end{center}

\vspace*{-0.3in} 
The results of our numerical study are presented in 
Figure~\ref{fig:CPX}, where we have plotted distributions in the 
invariant mass of the leading jet (identified as a Higgs jet) by the 
algorithm of Section~3. In every panel, the unshaded histograms 
represent the signal and the shaded histograms represent the $t\bar{t}$ 
background, which is the dominant SM background. Each row of panels 
corresponds to one of the benchmark points of Table~\ref{tab:BP1234}, as 
marked on the corresponding panels. The left (right) column corresponds 
to studies with a single (double) $b$-tag. To compensate for the lower 
efficiencies in the latter case, we use a higher luminosity of 
30~fb$^{-1}$ instead of the 10~fb$^{-1}$ used for the former.

The results for BP-1, BP-2 and BP-4 are very similar: in each case with 
a single $b$-tag we get a clear peak corresponding to the $h^0_1$ 
resonance. Taking the $t\bar{t}$ background alone, and the single bin of 
interest, this would be a deviation from the SM prediction at the level 
of around $10\sigma$ in each case --- which means that a discovery can 
be claimed in the very early stages of the 14~TeV LHC run. A similar 
statement can be made for the double $b$-tags with our higher luminosity 
assumption. For the benchmark point BP-3, with a single $b$-tag, a large 
and broad excess region can be found, with a small peak corresponding to 
$h^0_2$. On the other hand, for a double $b$-tag, the same benchmark 
point reveals rather disappointing results, with the signal hardly 
distinguishable from fluctuations in the $t\bar{t}$ histogram.

On the whole, therefore, we can conclude that prospects for detecting 
the lightest Higgs boson of the CP-violating MSSM by tagging `fat' jets 
are rather bright. The single $b$-tag alone gives very clear resonant 
peaks in the invariant mass distribution of the `fat' jets, and, if the 
parameters are favourable, one can expect confirming signals using the 
double $b$-tag method as well. It is only fair, however, to mention that 
our estimation of the background is somewhat crude and is limited to the 
$t\bar{t}$ signal at the leading order (LO). We know, of course, that 
addition of next-to-leading order (NLO) corrections to the $t\bar{t}$ 
cross section can increase the cross section by a factor of 2 or more. 
In that case, the background indicated by the shaded histogram in the 
figure will grow by a corresponding factor and the resonant peaks of 
the signal may not stand so tall above background as they appear in 
Figure~\ref{fig:CPX}. However, this is no reason to be disheartened, for 
($a$) the Higgs boson production modes will also receive comparable 
enhancements when NLO corrections are added, and ($b$) what will be 
observed will be the sum of signal and background events, and if this is 
compared with the random fluctuation in the background, the single 
$b$-tag signal with 10~fb$^{-1}$ of data will still be at a level of 
more than 5$\sigma$, which is all that we require for a discovery.

Before we end this section it is appropriate to comment on other 
backgrounds, apart from those due to $t\bar{t}$ production. The standard 
cuts devised for SUSY searches and applied to this signal will 
effectively remove backgrounds arising from the production of 
electroweak bosons, as may be guessed from the complete absence of peaks 
in the vicinity of $M_W$ and $M_Z$ in Figure~\ref{fig:CPX} --- unless, 
indeed, a light Higgs boson happens to lie just there. However, there 
will be a pure QCD background, which, despite drastic reduction by the 
same cuts and a very small mistagging probability, may still be expected 
to contribute something to the background because the initial cross 
section is many orders of magnitude larger than the signal. In this 
work, we have not attempted any detailed estimate of the QCD background 
or the mistagging probability, but a rough estimate based on the results 
of Ref.~\cite{CMSSusy} shows that the QCD background will not be more 
than a few femtobarns, whereas our signal, even after application of all 
cuts is around 100~fb. Another possibility is that of backgrounds 
arising from the production and decay of supersymmetric particles, which 
we have not estimated, but surely forms some fraction of what we have 
identified as the signal. Once again, we have not attempted a detailed 
study of these, but we can guess from the fairly tall and sharp 
resonances that we seem to be predicting for the light Higgs states, 
that these backgrounds will not really prove a difficulty when the 
experimental data are available.

\vspace*{-0.2in}
\section{Beyond the CPX Scenario}

\vspace*{-0.2in} We have seen in the previous section that one can use 
jet substructure analyses very effectively to probe hitherto-invisible 
Higgs bosons in the CPX scenario, as exemplified though our choice of 
four benchmark points. However, one can now ask whether a similar 
analysis can be used in a more general context than the CPX scenario, 
which itself is a benchmark created to showcase a `blind spot' in 
previous collider searches for the CP-violating MSSM. A detailed answer 
of this question would require a detailed scan of the parameter space to 
determine the exact extent of this collider-invisible region -- a 
lengthy and tedious business, given the large number of parameters which 
come into play when we allow the MSSM to become CP-violating. However, a 
partial answer, at least, can be sought by again picking up a couple of 
benchmark points, BP-5 and BP-6, which are do not conform to the CPX 
parametrisation, but are nevertheless part of the `blind spot', i.e. 
invisible to collider searches using conventional strategies.

\footnotesize
\begin{table}[htb]
\centering 
\begin{tabular}{c|cccccc|ccc|rc}
\hline
BP & $M_1$ & $M_2$ & $\tan\beta$ & $M(H^\pm)$ & $\mu$ & $\Phi_{\rm CP}$ &
$M(h^0_1)$ & $M(h^0_2)$ & $M(h^0_3)$ & 
$M(\tilde{\chi}_1^0)$ & $M(\tilde{\chi}_1^\pm)$ \\ 
\hline\hline
5 & 100 & 200 &   8 & 130 & 2400 & $\pi/2$   & 37.2 & 110.4 & 132.5 &  99.8 & 199.1 \\ [-2mm]
6 & 150 & 400 & 11 & 135 & 2000 & $3\pi/4$ & 56.8 & 117.3 & 127.7 & 149.8 & 398.7 \\
\hline
\end{tabular}
\def\baselinestretch{1.1}
\caption{\footnotesize Benchmark points lying outside the CPX scenario, 
but within the `blind spot' for collider searches. Free parameters not 
explicitly given in the table are taken identical with those in the CPX 
scenario of the previous section. All masses are in units of GeV.}
\def\baselinestretch{1.25}
\label{tab:BP56}
\end{table}
\normalsize 

As before, we exhibit our choice of benchmark points in 
Table~\ref{tab:BP56}, where, in addition to the free parameters, we 
display part of the mass spectrum.  The choice of BP-5 is dictated by a 
desire to make the lightest Higgs boson $h^0_1$ as light as possible -- 
and here it can be as low as about 37~GeV. The mass spectrum for BP-6 is 
rather similar to that of BP-3. The crucial difference here lies in the 
branching ratios exhibited in Table~\ref{tab:BR56}, where we note that 
the $\tilde{\chi}_2^0$ can decay with a reasonable branching ratio into 
the $ \tilde{\chi}_1^0$ plus any one of the three Higgs states $h^0_1$, 
$h^0_2$ and $h^0_3$. In this case, we could perhaps have comparable 
numbers of all the three Higgs states produced and look for all of them 
together using our tagging algorithm.

\footnotesize
\begin{table}[htb]
\centering 
\begin{tabular}{l|cc}
\hline
Process &  BP-5  & BP-6   \\ \hline\hline
$\tilde{\chi}_2^0 \to \tilde{\chi}_1^0 + h_1^0$ & 99.7 & 61.9   \\ [-1mm]
$\tilde{\chi}_2^0 \to \tilde{\chi}_1^0 + h_2^0$ & --     & 23.8  \\ [-1mm]
$\tilde{\chi}_2^0 \to \tilde{\chi}_1^0 + h_3^0$ & --     & 13.9 \\ \hline
$h^0_3 \to h_1^0 + h_1^0$ & 72.5  & 21.2 \\ [-1mm]
$h^0_2 \to h_1^0 + h_1^0$ & 88.4  & 30.5 \\ \hline
$h^0_3 \to b + \bar{b}$      & 23.5  &  71.3 \\ [-1mm]
$h^0_2 \to b + \bar{b}$      & 10.5  &  63 .0 \\ [-1mm]
$h^0_1 \to b + \bar{b}$      & 92.4  &   91.7 \\ \hline
$\tilde{t}_2 \to \tilde{t}_1 + h^0_1$ & 0.09 & 0.06  \\ [-2mm]
$\tilde{t}_2 \to \tilde{t}_1 + h^0_2$ & 20.2 & 17.5  \\ [-2mm]
$\tilde{t}_2 \to \tilde{t}_1 + h^0_3$ & 36.8 & 34.8  \\
\hline
\end{tabular}
\def\baselinestretch{1.1}
\caption{\footnotesize Some important branching ratios (per cent) for 
the four benchmark points of Table~\ref{tab:BP56}. Blank entries 
indicate that the corresponding decay is kinematically disallowed.}
\def\baselinestretch{1.25}
\label{tab:BR56}
\end{table}
\normalsize 

\begin{center}
\begin{figure}[h] 
\center{\includegraphics[width=0.65\textwidth]{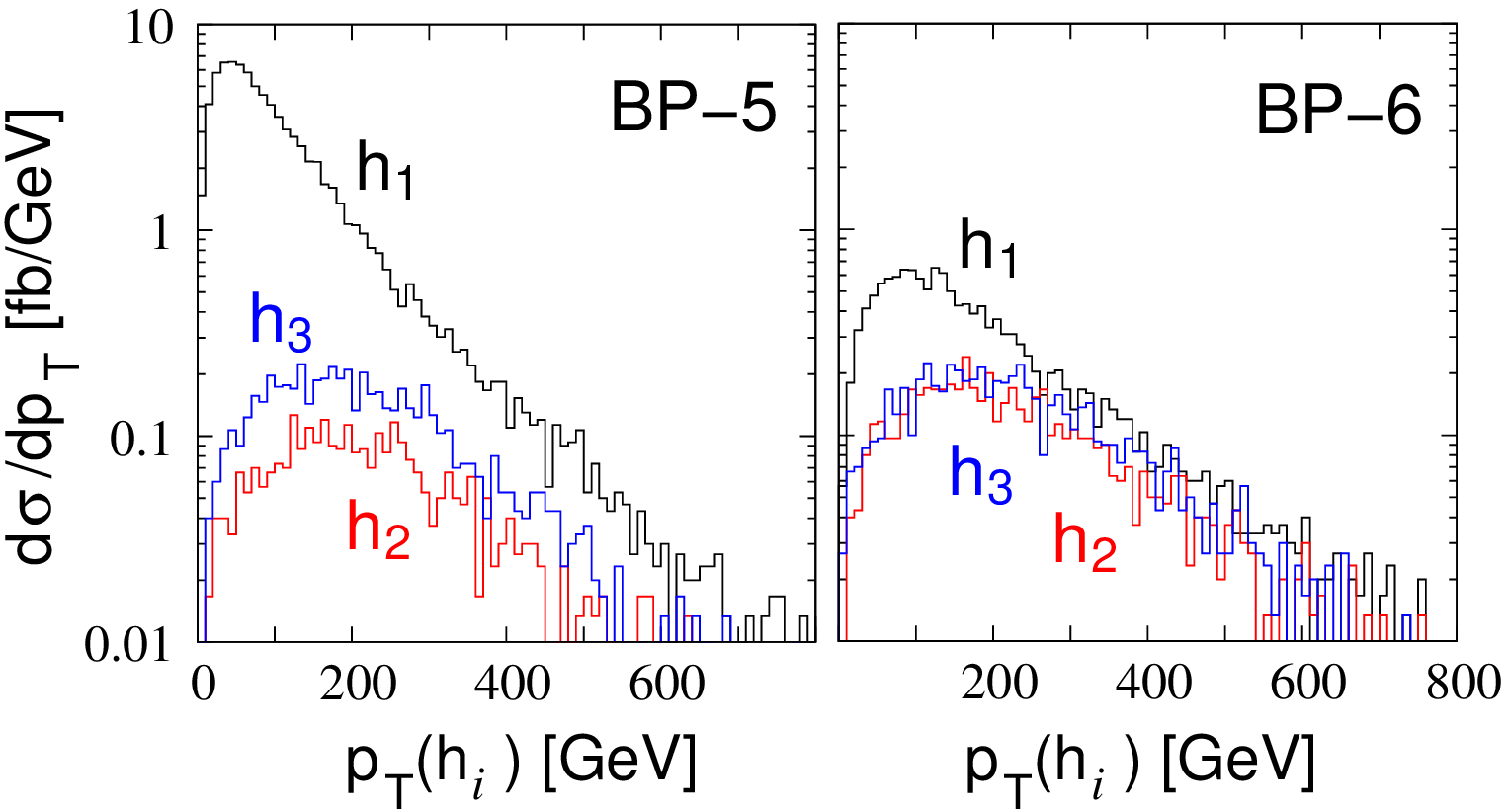}} 
\def\baselinestretch{1.1}
\caption{\footnotesize Distributions in transverse momentum $p_T(h_i)$ 
for the three Higgs bosons $h^0_i$ ($i = 1,2,3$) at the 14-TeV LHC.  
Black, red and blue histograms correspond, respectively, to the $h^0_1$, 
$h^0_2$ and $h^0_3$ states. The panels in this plot are marked BP-5 and 
BP-6, corresponding to the parameter choices of Table~\ref{tab:BP56}. No 
kinematic cuts were applied to generate these distributions.}
\def\baselinestretch{1.25}
\label{fig:pTdist2}
\end{figure}
\end{center}

Having made this choice, we once again start by investigating the 
$p_T$ distribution of the Higgs states to get a crude impression of how 
effective the Higgs tagging algorithm can be. Our results are presented 
in Figure~\ref{fig:pTdist2}, which resembles Figure~\ref{fig:pTdist1} 
closely, except that the panels correspond to BP-5 and BP-6 rather than 
BP-1 to BP-4. The histograms in black, red and blue correspond to the 
$h^0_1$, $h^0_2$ and $h^0_3$ states respectively. The left (right) panel 
corresponds to BP-5 (BP-6) as marked. The qualitative features of these 
two plots are very different: for BP-5, production of $h^0_1$ bosons is 
overwhelmingly dominant over production of $h^0_2$ and $h^0_3$, whereas 
the numbers are more comparable for BP-6. The reason for this lies less 
in the kinematics than in the dominant branching ratio of 
$\tilde{\chi}_2^0$ decays to the $h^0_1$ in the BP-5 case, as exhibited 
in Table~\ref{tab:BR56} below.

\vspace*{-0.2in}

\vspace*{-0.1in} Once again, it is interesting to compare 
Figure~\ref{fig:pTdist2} with the efficiency plots in 
Figure~\ref{fig:efficiency}. Concentrating on BP-5 for the moment, let 
us note that for a $h^0_1$ mass as low as 37~GeV, reasonable 
Higgs-tagging efficiencies can be obtained in the $p_T$ range $150 - 
400$~GeV, which allows us to include a little more of the peak region of 
the $p_T$ histogram than was the case for, say, BP-1. Thus we may expect 
a somewhat taller resonance for the $h^0_1$ in this case than was found 
for BP-1 and similar cases. In this case, the number of $h^0_2$ and 
$h^0_3$ produced is too small to expect any signals over the substantial 
$t\bar{t}$ background. If we carry out a similar analysis for the case 
of BP-6, we will find that all three Higgs bosons have comparable 
chances of producing a resonance in tagged `fat' jets, though even here 
the $h^0_1$ will have a slight edge over the others due to its lighter 
mass and greater tagging efficiency.

The above arguments enable us to make some crude guesses about the 
possible results, but it is necessary to carry out the full Monte Carlo 
simulation as in the previous section before we can say anything more 
definitive. In order to do this, we use the same machinery as described 
in the previous section, with the identical set of cuts and efficiency 
assumptions.

\vspace*{-0.2in}
\begin{center}
\begin{figure}[h] 
\center{\includegraphics[width=0.65\textwidth]{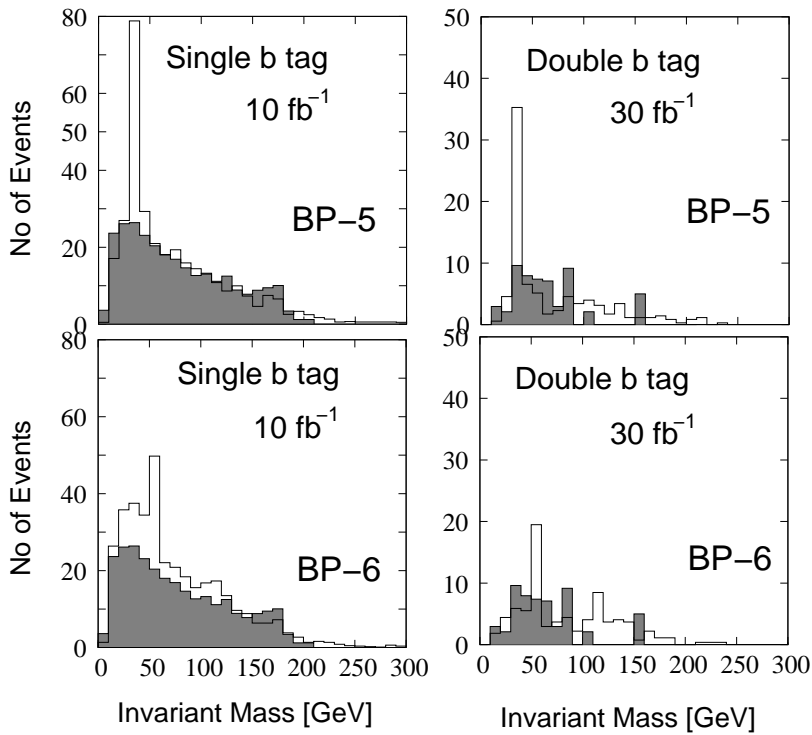}}
\def\baselinestretch{1.1}
\caption{\footnotesize Bin-wise invariant mass distribution of the 
leading jet with single (double) $b$-tags in 10 (30) fb$^{-1}$ of data 
at the 14-TeV LHC for the benchmark points BP-5 and BP-6 (marked on the 
respective panels). Note the clear $h^0_1$ resonance for BP-5 with both 
single and double $b$-tags. For BP-6 we find small resonances 
corresponding to the $h^0_1$ and $h^0_2$. The shaded histogram 
represents the $t\bar{t}$ background. }
\def\baselinestretch{1.25}
\label{fig:CPY}
\end{figure}
\end{center}

\vspace*{-0.4in} The results of our numerical simulations are displayed 
in Figure~\ref{fig:CPY}, where, as in Figure~\ref{fig:CPX}, we plot 
distributions in the invariant mass of the leading jet (identified as a 
Higgs jet by the algorithm of Section~3). The upper row of panels 
corresponds to the benchmark point BP-5 of Table~\ref{tab:BP56}, as 
marked on the panels, while the lower row corresponds to the benchmark 
point BP-6. As before, the left column corresponds to studies with a 
single $b$-tag with 10~fb$^{-1}$, while the right column presents the 
results with a double $b$-tag and 30~fb$^{-1}$ of data.  As in 
Figure~\ref{fig:CPX}, the unshaded histograms represent the signal while 
the shaded histograms represent the $t\bar{t}$ background.

These results are more-or-less as expected from the previous discussion, 
where we took the $p_T$ distribution and the efficiency plot of 
Figure~\ref{fig:efficiency} jointly into consideration. For the 
benchmark point BP-5, with a 37~GeV Higgs state, we get tall resonances 
in the `fat' jet invariant mass. Here the single $b$-tag method would 
produce a signal even for 1~fb$^{-1}$ of data. Of all the benchmark 
points considered, this is definitely the best signal we get, and thus, 
if we are lucky, even the first few months of running of the LHC at 
14~TeV would throw up a light Higgs discovery. On the other hand, if we 
consider the other benchmark point BP-6, the signal is much weaker, but 
it seems just possible to detect the $h^0_1$ as well as an $h^0_2$ 
state. The evidence for the latter, as evinced in Figure~\ref{fig:CPY}, 
is not very strong, but would certainly improve as more data are 
collected.  Discovery of a second Higgs boson would go a long way 
towards establishing a CP-violating Higgs sector, and more generally, a 
BSM Higgs sector.

\vspace*{-0.2in}
\section{Summary and Outlook}

\vspace*{-0.2in} In this article, we have presented arguments to show 
that the existence of light Higgs scalars with masses below 100~GeV is 
by no means ruled out by existing data, except within the very 
restrictive assumptions made in the SM and in the MSSM without CP 
violation.  As an example of a model where relaxation of these 
assumptions permits the existence of light Higgs bosons, we have chosen 
the CP-violating version of the MSSM, and argued that this is by no 
means an exotic model, but a very natural version of a supersymmetric 
model. In this model, we have picked on a set of parameters known as the 
CPX scenario, which is known to lie in a `blind spot' of all collider 
searches till the present -- including the LEP-1 and LEP-2, the Tevatron 
and the just-concluded runs of the LHC. To keep the discussion focussed, 
we have chosen four benchmark points in the CP-violating MSSM, all of 
which lie within the CPX scenario, and verified that these all 
correspond to light Higgs bosons which would be completely invisible to 
all the above-mentioned collider searches. Of course, except for our 
specific choice of the benchmark points, all of this is simply a 
reiteration of facts or results available in the literature.

The novel feature of our work is the suggestion that the `blind spot' of 
previous collider searches may actually be explored at the LHC using 
recently-developed Higgs-tagging techniques using jet substructure. Once 
again, such techniques have been described in the literature earlier, 
and even applied to study Higgs bosons in the CP-conserving MSSM, though 
with somewhat modest results\cite{GhoGuSen}.  In our work, we simply 
follow the original algorithm for Higgs tagging, with some small 
modifications to the case of interest, as described in Section 3. The 
results of our study are showcased in Figure~\ref{fig:efficiency}, where 
the efficiency contours in the plane of Higgs $p_T$ and Higgs mass 
clearly show that the method is much more efficient for light Higgs 
bosons than for Higgs bosons in the range permitted by the SM and 
CP-conserving MSSM. When we apply it to the benchmark points chosen 
within the CPX scenario, we find that clearly-identifiable resonances 
corresponding to the lightest Higgs boson appear in the invariant mass 
distribution of all jets which pass through the Higgs tagging algorithm. 
For some points, one can even discern hints of a second Higgs boson. 
These encouraging results appear to be found whether we tag one or both 
of the $b$ sub-jets resulting from the Higgs decays.

Emboldened by our findings, we have extended our study beyond the CPX 
scenario to other parts of the `blind' region, where we have selected 
another two benchmark points, with distinct features in the mass spectra 
and branching ratios. Here again, we have found very clear resonances, 
indicating that the same technique can be a powerful probe of a large 
part, if not all, of the hitherto `blind spot' of collider searches in 
the CP-violating MSSM, over and above the CPX region.

The purpose of this work was to establish that the Higgs-tagging 
algorithm can be used effectively to probe the `blind spot' of the 
CP-violating MSSM, and we believe that has been adequately established 
by our getting positive results at all the six benchmark points chosen. 
A more detailed mapping of the `blind spot' is in progress\cite{selves}. 
In fact, we have already remarked that this method can also probe other 
models with light Higgs bosons, for the Higgs-tagging algorithm works so 
long as these Higgs bosons are boosted and decay to $b\bar{b}$ pairs.

Eventually, we hope that this powerful technique will be taken up by the 
experimental collaborations and applied to real data, instead of Monte 
Carlo simulations as was done in this exploratory study. The most 
exciting possibility, of course, would be if experimental searches could 
use this technique to actually find a light Higgs resonance, which has 
been missed in all collider searches so far. If not, we will have to 
settle, as usual, for a more constrained parameter space for the model 
in question. In either case, the present technique could be the key to 
accessing the region which previous searches could not, and that alone 
represents a modest degree of progress.

\bigskip

\small
\baselineskip 16pt {\bf Acknowledgement} \\
\vskip -15pt
The work of BB is supported by World Premier International Research 
Center Initiative (WPI Initiative), MEXT, Japan. DKG acknowledges 
partial support from the Department of Science and Technology, India, 
under the grant SR/S2/HEP-12/2006. BB, AC and DKG are grateful to, 
respectively, the IACS (Kolkata, India), the TIFR (Mumbai, India) and 
the KIAS (Seoul, Korea), for hospitality while part of this work was 
being done. AC thanks Subir Sarkar for programming help and guidance, 
while SR acknowledges useful discussions with Tuhin S. Roy.

\newpage            

\end{document}